\documentclass[dvips,12pt,a4paper]{article}
\usepackage{epsfig}
\usepackage{a4}
\usepackage{amsmath}
\usepackage{latexsym}

\def\lsim{\mathrel{\rlap{\raise 2.5pt \hbox{$<$}}\lower 2.5pt
\hbox{$\sim$}}}
\def\gsim{\mathrel{\rlap{\raise 2.5pt \hbox{$>$}}\lower 2.5pt
\hbox{$\sim$}}}

\def\GeV{{\rm GeV}}

\begin{document}
\thispagestyle{empty}
\phantom{AAA}
\vspace*{-20mm}
\begin{center}
{\bf{\Large Particle Mass Limits in  
Minimal and \\ Nonminimal Supersymmetric Models
\footnote{Invited talk given at the Fifth Workshop on High Energy Physics
Phenomenology, Inter-University Centre for Astronomy and
Astrophysics, Pune, India, January 12 - 26, 1998 }}}
\vskip 1.0cm
P. N. Pandita\\ 
Department of Physics, North Eastern Hill University\\
Umshing-Mawkynroh, Shillong 793 022, India\\[-2mm]
\end{center}
\vskip 1.0cm
\begin{abstract}
A review of the Higgs and neutralino sector of supersymmetric models
is presented. This includes the upper limit on the
mass of the lightest Higgs boson in the 
Minimal Supersymmetric Standard
Model, as well as models based on the Standard Model
gauge group $SU(2)_L \times U(1)_Y$ with extended Higgs sectors.
We then discuss the Higgs sector of left-right supersymmetric
models, which conserve R-parity as a consequence of gauge invariance,
and present a calculable upper bound on the mass of the
lightest Higgs boson in these models. We also discuss the
neutralino sector of general
supersymmetric models based on the SM gauge group.
We show that, as a consequence of gauge coupling unification,
an  upper bound on the mass of the lightest neutralino
as a function of the gluino mass can be obtained.
\end{abstract}
\newpage
\section{Introduction}
The Standard Model (SM) of elementary particle physics, which is
extremely  successful
in describing the experimental data,
is based on two fundamental principles, i.e. gauge invariance and Higgs
mechanism. From recent experiments it is clear that strong and
electroweak interactions are described by an $SU(3)_C \times SU(2)_L
\times U(1)_Y$ gauge theory. On the other hand, little is known
about the mechanism of electroweak
gauge symmetry breaking.  In the Standard Model, the
electroweak $SU(2)_L\times U(1)_Y$ gauge symmetry is broken 
through the Higgs mechanism via the 
vacuum expectation value
(VEV) of the neutral component of the Higgs doublet
\cite{Gunion:1989we}, leaving behind a remanant in the form 
of an elementary scalar particle, the Higgs boson, which has  so far not
been observed.
Apart from the fact that the VEV is an arbitrary
parameter in the SM, the mass parameter of the Higgs field suffers
from quadratic divergences, making the weak scale unstable under
radiative corrections~\cite{Gunion:1989we}. 

One of the central problems of particle physics is, then,
to understand how the electroweak scale associated with the mass 
of the W boson is
generated, and why is it so small as compared to the Planck scale
associated with the Newton's constant.
Supersymmetry \cite{Haber:1985rc,Nilles:1984ex,Nath:1983fp}
is at present the only known
framework in which the weak scale is stable under radiative
corrections, although it does not explain
how such a small scale arises in the first place. As such,
considerable importance attaches to the study of supersymmetric
models, especially the Minimal Supersymmetric Standard Model (MSSM),
based on the gauge group $SU(3)_C \times SU(2)_L \times U(1)_Y$, 
with two Higgs doublet superfields $(H_1, H_2)$
with opposite hypercharges: $Y(H_1) = -1$, $Y(H_2) = +1$, so as to
generate masses for up- and down-type
quarks (and leptons), and to cancel gauge anomalies.
In general, supersymmetric extensions of SM have extended Higgs sectors
leading to a rich penomenology of Higgs bosons.

In this talk I will discuss the Higgs sector of supersymmetric models.
This will include the Minimal Supersymmetric Standard Model,
as well as models having extended Higgs sectors, based
on the SM gauge group. I will then explore supersymmetric
models based on extended gauge groups, e.g. the left-right
gauge group $SU(2)_L \times SU(2)_R \times U(1)_{B-L}$, 
pointing out the relevance of extended gauge symmetries
in the context of supersymmetric models,
and discuss the Higgs sector of these models.

In supersymmetric gauge theories, each fermion and boson of the
Standard Model is
accompanied  by its supersymmetric partner, transforming in an
identical manner under the gauge group.  In
supersymmetric theories with R-parity conservation,
the lightest
supersymmetric particle (LSP) is expected to be the lightest
neutralino, which is the lightest mixture of the fermionic
partners of the neutral Higgs and neutral electroweak gauge
bosons. The lightest neutralino, being the LSP, 
is the end product of any process involving supersymmetric particles
in the final state. In this talk, I will also discuss the 
neutralino sector of the general supersymmetric models
based on the SM gauge group.

\section{The Higgs Sector of the Minimal Supersymmetric Standard Model}

The Higgs sector of the Minimal Supersymmetric Standard Model,
based on the gauge group $SU(3)_C \times SU(2)_L \times U(1)_Y$,
contains  two Higgs doublet superfields $(H_1, H_2)$
with opposite hypercharges: $Y(H_1) = -1$, $Y(H_2) = +1$, so as to
generate masses for up- and down-type
quarks (and leptons), and to cancel gauge anomalies.
The tree level scalar potential of Higgs bosons in
MSSM can be wriitten as 
\begin{eqnarray}
V_H&=&
m_1^2 \mid H_1\mid^2 + m_2^2 \mid H_2\mid^2
- m_3^2 (H_1 \cdot  H_2+ {\rm h.c.} ) \nonumber \\
& + & \frac {g^2+g^{\prime 2}}{8}
(\mid H_1\mid^2 - \mid H_2\mid^2)^2
+ \frac{1}{2} g^2 \mid H_1^*H_2\mid^2,
\label{eq:vhiggs}
\end{eqnarray}
where $g$ and $g^{\prime}$ are the $SU(2)_L$ and $U(1)_Y$ gauge couplings, 
respectively. 
We note that, as a consequence of gauge invariance and supersymmetry,
the quartic couplings of Higgs bosons in MSSM 
are fixed in terms of electroweak gauge couplings.
After spontaneous symmetry breaking induced by the neutral
components of $H_1$ and $H_2$ obtaining vacuum
expectation values, $\langle H_1\rangle = v_1$,
$\langle H_2\rangle = v_2$, $\tan\beta = v_2/v_1$,
the MSSM contains two neutral $CP$-even
($h^0$, $H^0$), one neutral
$CP$-odd ($A$), and two charged ($H^{\pm}$) Higgs
bosons~\cite{Gunion:1989we}.
Although gauge invariance and supersymmetry fix the quartic couplings
of the Higgs bosons in MSSM in terms of $SU(2)_L$ and $U(1)_Y$
gauge couplings, there still
remain two independent parameters which describe the Higgs sector
of the MSSM. These are usually chosen to be $\tan\beta$ and
$m_A$, the mass of the $CP$-odd Higgs boson. All the Higgs masses
and the Higgs couplings in MSSM can be described (at tree level)
in terms of these two parameters. From 
(\ref{eq:vhiggs}) it follows  
that the lightest CP-even neutral Higgs boson has a tree level
upper bound of $M_Z$ (the mass of $Z^0$ boson) on its mass
\cite{Inoue:1982ej,Flores:1983pr}.
However, radiative corrections 
\cite{Okada:1991vk,Ellis:1991nz,Haber:1991aw,Barbieri:1991ja}
weaken  this tree level upper bound. 
In the one-loop effective potential approximation,
the radiatively corrected squared mass matrix for the
$CP$-even Higgs bosons can be written as \cite{ERZ2}
\begin{eqnarray}
\label{eq:hmass}
{\mathcal M}^2
&=&  \left[ \begin{array}{cc}
m_A^2 \sin^2 \beta + m_Z^2 \cos^2\beta &
-(m_Z^2 + m_A^2) \sin\beta \cos \beta\\
-(m_Z^2 + m_A^2) \sin\beta \cos \beta &
m_A^2 \cos^2 \beta + m_Z^2 \sin^2\beta
\end{array} \right] \nonumber \\
& &
+ \frac{3 g^2}{16 \pi^{2} m_W^2}
\left[ \begin{array}{cc}
\Delta_{11} & \Delta_{12}\\
\Delta_{12} & \Delta_{22}
\end{array} \right],
\end{eqnarray}
where the second matrix represents the radiative
corrections.

The functions $\Delta_{ij}$ depend on,
besides the top- and bottom-quark masses, the Higgs
bilinear parameter $\mu$ in the super-potential,
the soft supersymmetry breaking trilinear couplings
($A_t$, $A_b$) and soft scalar masses ($m_Q$, $m_U$, $m_D$),
as well as $\tan\beta$.
We shall ignore the b-quark mass effects in
$\Delta_{ij}$ in the following, which is a reasonable
approximation for moderate values of $\tan\beta \leq 20-30$.
Furthermore, we shall assume, as is often done,
\begin{eqnarray}
A &\equiv& A_t = A_b, \nonumber \\
\tilde m &\equiv& m_Q = m_U = m_D. \label{2}
\end{eqnarray}
With these approximations we can write ($m_t$ is the top quark mass)
\cite{ERZ2}
\begin{eqnarray}
\Delta_{11} & = & \frac{m_t^4}{\sin^2\beta}
                \left(\frac{\mu(A+\mu\cot\beta)}
                {m_{\tilde t_1}^2 - m_{\tilde t_2}^2} \right)^2
                g(m_{\tilde t_1}^2, m_{\tilde t_2}^2), \label{3}\\
\Delta_{22} & = & \frac{m_t^4}{\sin^2\beta}
 \left(\log\frac{m_{\tilde t_1}^2 m_{\tilde t_2}^2}{m_t^4}
+  \frac{2A(A+\mu\cot\beta)}{m_{\tilde t_1}^2 - m_{\tilde t_2}^2}
   \log\frac{m_{\tilde t_1}^2}{m_{\tilde t_2}^2} \right ) \nonumber \\
 & + & \frac{m_t^4}{\sin^2\beta}
       \left(\frac{\mu(A+\mu\cot\beta)}
      {m_{\tilde t_1}^2 - m_{\tilde t_2}^2} \right)^2
     g(m_{\tilde t_1}^2, m_{\tilde t_2}^2), \label{4}\\
\Delta_{12} & = & \frac{m_t^4}{\sin^2\beta} \frac{\mu(A+\mu\cot\beta)}
     {m_{\tilde t_1}^2 - m_{\tilde t_2}^2}
      \left(\log\frac{m_{\tilde t_1}^2}{m_{\tilde t_2}^2}
       +  \frac{A(A+\mu\cot\beta)}{m_{\tilde t_1}^2 - m_{\tilde t_2}^2}
       g(m_{\tilde t_1}^2, m_{\tilde t_2}^2) \right), \label{5}
\end{eqnarray}
where $m_{\tilde t_1}^2$ and $m_{\tilde t_2}^2$ are squared stop masses,
and $g(m_{\tilde t_1}^2, m_{\tilde t_2}^2)$ is a function of 
stop masses, given by
(we have ignored the small $D$-term contributions to the stop masses)
\begin{eqnarray}
m_{\tilde t_{1,2}}^2   &=&  m_t^2 + \tilde m^2 \pm
m_t(A + \mu\cot\beta),\label{6}\\
g(m_{\tilde t_1}^2, m_{\tilde t_2}^2) &=&  2 -
\frac{m_{\tilde t_1}^2 + m_{\tilde t_2}^2}{m_{\tilde t_1}^2 -
m_{\tilde t_2}^2} \log\frac{m_{\tilde t_1}^2}{m_{\tilde t_2}^2}.
\label{7}
\end{eqnarray}
The one-loop radiatively corrected masses ($m_{h}$, $m_{H}$;
$m_{h} <  m_{H}$) of the $CP$-even Higgs bosons ($h^0$, $H^0$)
can be obtained by diagonalizing the $2 \times 2$ mass matrix
(\ref{eq:hmass}). The radiative corrections are, in general, positive,
and they shift the mass of the lightest Higgs boson upwards from
its tree-level value.
We show in Fig.(1) the resulting masses of
the CP-even Higgs bosons,
$m_h$ and $m_H$, as well as the charged Higgs boson mass,
as a function of $m_A$ for two different
values of $\tan\beta = 1.5, 30$.
With a wider range of parameter values,
or when the squark mass scale
is taken to be smaller,
the dependence on $\mu$ and $\tan\beta$ can be more
dramatic \cite{KOP1, KOP2, KOP3}.

Although radiative corrections can be appreciable,
these depend only
logarithmically on the SUSY breaking scale, and are, therefore, under
control. In particular, the mass of the lightest neutral
scalar is bounded from above:
\begin{equation}
\label{eq:MSSM1}
m_h^2 \leq M_Z^2 \cos^2(2\beta) + \epsilon(m_t, m_{\tilde t_1},
m_{\tilde t_2}, A_t, \mu, \cdot \cdot \cdot),
\end{equation}
where $\epsilon$ parameterizes the effect of the radiative corrections
described above. Note that $\epsilon$ is approximately independent of
$\tan\beta$; for large $m_A$, $m_t=175$ GeV and $m_{\tilde t_{1,2}}=1$ TeV it
amounts to about $0.9 M_W^2 \ (1.6 M_W^2)$ for no (maximal) stop mixing.
It is important to note that the bound (\ref{eq:MSSM1}) can {\em only} be
saturated for large $m_A$. This rsults in an upper bound of 
about $125-135$ GeV on the
one-loop radiatively corrected mass of the lightest
Higgs boson of MSSM \cite{Rosiek:1995dy}.

The Higgs mass falls rapidly at small values of $\tan\beta$.
Since the LEP experiments are obtaining
lower bounds on the mass of the lightest Higgs boson, they are
beginning to rule out significant parts of the small-$\tan\beta$
parameter space, depending on the model assumptions.
For $\tan\beta>1$, ALEPH finds $m_h>62.5~\GeV$ at 95\% C.L.\ \cite{ALEPH}.
[For a recent discussion on how the lower allowed value of $\tan\beta$
depends on some of the model parameters, see Ref.~\cite{CCPW}.]

The two-loop corrections to the lightest higgs mass are typically
$\mathcal{O}$$(20\%)$ of the one-loop corrections, and are negative.
For the dominant two-loop radiative corrections
to the Higgs sector of MSSM, see,  e.g.\ \cite{Carena1, Carena2}.

\section{The Higgs Sector of the Non-Minimal Supersymmetric Standard Model}

If we concentrate on the Higgs sector, the MSSM is special because
the Higgs self-couplings at the tree level are completely determined
by gauge couplings. Although the MSSM is the simplest, and, thus,
the  most widely, studied model, there 
are several viable extensions of the supersymmetric version of the SM. 
A simplest extension of the Higgs sector
of the MSSM is to postulate the existence of a 
$SU(2)_L \times U(1)_Y$ Higgs singlet superfield $N$ in the spectrum
~\cite{NSW}.
This model, referred to as the Non-Minimal Supersymmetric Standard
Model (NMSSM),  does not destroy the unification of coupling constants
achieved in the MSSM, since the new light particles do not carry
SM quantum numbers.

Even if we restrict ourselves to purely cubic terms in the
superpotential $W$, gauge symmetry allows one to introduce two
different Higgs self--couplings:
\begin{equation}
\label{eq:NMSSM1}
W_{\rm Higgs} = \lambda N H_1 H_2 - \frac {k}{3} N^3,
\end{equation}
where we have used the notation of Ref.~\cite{EGHRZ}. Together with the
corresponding soft breaking terms, there are six free parameters in the
Higgs sector, even after we fix the sum of the squares of the VEV's
of the $SU(2)$ doublets to reproduce the known mass of the $Z$ boson.
Moreover, the spectrum now contains three neutral CP--even fields
$H_i$ and two CP--odd fields $A_i$ in addition to the charged Higgs field
$H^\pm$.

Because of the presence of the trilinear coupling proportional 
$\lambda$ in the superpotential (\ref{eq:NMSSM1}), the tree-level
Higgs-boson self-coupling in the NMSSM 
depends on $\lambda$ as well as the gauge  couplings.
Nevertheless, one can still 
derive~\cite{Drees:1989fc,Binetruy:1992mk,Moroi:1992zk,Pandita:1993hx,
Pandita:1993tg,Elliott:1993ex,Ellwanger:1993hn} an upper bound
on the mass of the lightest CP-even Higgs boson  of the NMSSM. Including
radiative corrections, one has
\begin{equation}
\label{eq:NMSSM2}
m^2_{H_1} \leq M_Z^2 \cos^2(2\beta) + \frac {2 \lambda^2 M_W^2} {g^2}
\sin^2(2\beta) + \epsilon,
\end{equation}
where $\epsilon$ parametrizes the effect of radiative corrections, which 
are similar in nature to the corresponding corrections in the MSSM.
Because of the presence of the term proportional to
the coupling $\lambda$ in (\ref{eq:NMSSM2}), 
no definite upper bound on the
mass of the lightest CP-even Higgs boson in NMSSM can be given 
unless a further assumption on the strength of this
coupling is made. If we require all dimensionless coupling constants
to remain perturbative upto the GUT scale, we can
calculate the the upper bound on the lightest CP-even
Higgs-boson mass. 
The resulting upper bound is shown 
in Fig.(2), and is compared with the corresponding bound
in the MSSM~\cite{Pandita:1993hx,Pandita:1993tg,AP1,KOT1}.
The top-quark-mass
dependence of the upper bound
is not significant compared to the MSSM 
case because the maximally allowed value of $\lambda$ 
is larger(smaller) for a smaller (larger) top mass.

One can study the implications of introducing higher
dimensional Higgs representations on the
upper bound for the mass of the lightest 
Higgs boson in supersymmetric models.
Because of the presence of the additional trilinear
Yukawa couplings, a tight constraint on the mass of the lightest
Hiss boson need not {\em a priori} hold in such 
extensions of MSSM
based on the gauge group $SU(2)_L \times U(1)_Y$ with an
extended Higgs sector. Nevertheless, it has been shown that the upper
bound on the lightest Higgs boson mass in these models depends only on
the weak scale and dimensionless coupling constants (and only
logarithmically on the SUSY breaking scale), and is calculable if all
the couplings remain perturbative below some scale $\Lambda$
\cite{ Kane:1993kq,Espinosa:1993hp}.
This upper
bound can vary between 150 GeV to 165 GeV depending on the Higgs
structure of the underlying supersymmetric model. Thus, nonobservation
of a light Higgs boson below this upper bound will rule out an entire
class of supersymmetric models based on the gauge group $SU(2)_L
\times U(1)_Y$.

\section{Supersymmetric Models with Extended gauge Groups}

The existence of upper bound on the lightest Higgs boson mass in MSSM
(with arbitrary Higgs sectors) has been investigated in a situation
where the underlying supersymmetric model respects baryon ($B$) and
lepton ($L$) number conservation. However, it is well known that gauge
invariance, supersymmetry and renormalizibility allow $B$ and $L$
violating terms in the superpotentioal of the MSSM
\cite{Weinberg:1982wj,Sakai:1982pk}.
The strength of these lepton and baryon number violating terms is,
however, severely limited by phenomenological
\cite{Zwirner:1984is,Hall:1984id,Lee:1984tn,Dawson:1985vr,Barbieri:1986ty,Dimopoulos:1988jw,Barger:1989rk,Dreiner:1991pe,Smirnov:1996ey}
and cosmological
\cite{Campbell:1991fa,Fischler:1991gn} constraints.
Indeed, unless the strength of
baryon-number-violating term is less than $10^{-13}$,
it will lead to contradiction with the present
lower limits on the lifetime of the proton. The usual strategy to
prevent the appearance of $B$ and $L$ violating couplings in MSSM is
to invoke a discrete $Z_2$ symmetry
\cite{Farrar:1978xj}
known as matter parity, or R-parity. The matter parity of 
each superfield may be defined as
\begin{equation}
(\text{matter} \text{ parity}) \equiv (-1)^{3 (B-L)} .
\label{eq:rparity}
\end{equation}
The multiplicative conservation of matter parity forbids all the
renormalizable $B$ and $L$ violating terms in the superpotential of
MSSM. Equivalently, the R-parity of any component {\em field} is
defined by $R_p = (-1)^{3(B-L)+2S}$, where $S$ is the spin of the
field. Since $(-1)^{2S}$ is conserved in any Lorentz-invariant
interaction, matter parity conservation and R-parity conservation are
equivalent. Conservation of R-parity  then immediately implies
that superpartners can be produced only in pairs, and that the
lightest supersymmetric particle (LSP) is absolutely stable.

Although the Minimal Supersymmetric Standard Model with R-parity
conservation can provide a description of nature which is consistent
with all known observations, the assumption of $R_p$ conservation
appears to be {\em ad hoc}, since it is not required for the internal
consistency of MSSM. Furthermore, all global symmetries, discrete or
continuous, could be violated by the Planck scale physics effects
\cite{Giddings:1988cx,Coleman:1988tj,Preskill:1990bm,Holman:1992us,Kamionkowski:1992mf}. The problem becomes acute for low energy
supersymmetric models, because $B$ and $L$ are no longer automatic
symmetries of the Lagrangian, as they are in the Standard  Model.
It is, therefore, more appealing to have an supersymmetric theory
where R-parity is related to a gauge symmetry, and its conservation is
automatic because of the invariance of the underlying theory
under this gauge symmetry. Fortunately, there is a compelling
scenario which does
automatically provide for exact R-parity conservation due to a deeper
principle. Indeed, $R_p$ conservation follows automatically in certain
theories with gauged $(B-L)$, as is suggested by the fact that matter
parity is simply a $Z_2$ subgroup of $(B-L)$. It has been noted by
several authors
\cite{Mohapatra:1986su,Font:1989ai,Martin:1992mq}
that if the gauge symmetry
of MSSM is extended to $SU(2)_L \times U(1)_{I_{3R}} \times
U(1)_{B-L}$, or $SU(2)_L \times SU(2)_R \times U(1)_{B-L}$, the theory
becomes automatically R-parity conserving. Such a left-right
supersymmetric theory (SLRM) solves the problems of explicit $B$ and
$L$ violation of MSSM, and has received much attention recently
\cite{Cvetic:1984su,Francis:1991pi,Frank:1993ku,Kuchimanchi:1993jg,Huitu:1994gf,Huitu:1994uv,Huitu:1995zm,Huitu:1997iy,Aulakh:1997ba}.
Of course left-right theories are also interesting in their own right,
for among other appealing features, they offer a simple and natural
explanation for the  smallness of neutrino mass through the
so called see-saw
mechanism \cite{Gell-Mann:1979,Yanagida:1979}.

Such a naturally R-parity conserving theory necessarily involves the
extension of the Standard Model gauge group, and since the extended
gauge symmetry has to be broken, it involves a ``new scale'', the
scale of left-right symmetry breaking, beyond the SUSY and $SU(2)_L
\times U(1)_Y$ breaking scales of MSSM. It is, therefore, important to
ask whether the upper bound on the lightest Higgs mass in naturally
R-parity conserving theories depends  on the scale of the breakdown of
the extended gauge group. We now consider the question of the
mass of the lightest Higgs boson in 
left-right supersymmetric models so as
to answer this question~\cite{HPP2}.

We begin by recalling some basic features of the left-right
supersymmetric models used in our study.
Further details can be found, e.g., in \cite{Huitu:1997iy, HPP2}.
The quark and lepton doublets are included in
$Q(2,1,1/3)$; $Q^c(1,2,-1/3)$; $L(2,1,-1)$; $L^c(1,2,1)$,
where $Q$ and $Q^c$
denote the left- and right-handed quark superfields
and similarly for the leptons $L$ and $L^c$.
Note that since left- and right-handed fermions are placed
symmetrically in
doublets, also the right-handed neutrinos are included.
The Higgs superfields consist of
$\Delta_L(3,1,-2)$; $\Delta_R(1,3,-2)$;
$\delta_L(3,1,2)$; $\delta_R(1,3,2)$;
$\Phi(2,2,0)$; $\chi(2,2,0)$.
The numbers in the parentheses denote the representation
content of the fields under the gauge group
$SU(2)_L\times SU(2)_R\times U(1)_{B-L}$.
The two $SU(2)_R$ Higgs triplet superfields
$\Delta_R (1,3,-2)$ and $\delta_R(1,3,2)$ with
opposite $(B-L)$ are necessary to break the
left-right symmetry spontaneously, and to cancel
triangle gauge anomalies due to the fermionic superpartners
of Higgs bosons.
The left-right model also contains the $SU(2)_L$ triplets
$\Delta_L$ and $\delta_L$ in order to make the Lagrangian fully
symmetric under the  $L\leftrightarrow R$ transformation,
although these are not needed phenomenologically for
the symmetry breaking or the see-saw 
mechanism~\cite{Gell-Mann:1979,Yanagida:1979}
for neutrino mass generation.
The two bi-doublet Higgs
superfields $\Phi$ and $\chi$ are required
in order to
break the $SU(2)_L\times U(1)_Y$
and to generate a non-trivial Kobayashi-Maskawa matrix.

The most general gauge invariant superpotential for the model can be written as
\begin{eqnarray}
W&=& h_{\phi Q}Q^T i\tau_2 \Phi Q^c +
h_{\chi Q}Q^T i\tau_2 \chi Q^c +
h_{\phi L}L^T i\tau_2 \Phi L^c
+ h_{\chi L}L^T i\tau_2 \chi L^c \nonumber\\
&&+h_{\delta_L} L^T i\tau_2 \delta_L L +
h_{\Delta_R} L^{cT} i\tau_2 \Delta_R L^c+
\mu_1 \text{Tr} (i\tau_2\Phi^T i\tau_2 \chi) +
\mu_1' \text{Tr} (i\tau_2\Phi^T i\tau_2 \Phi) \nonumber\\
&&+ \mu_1'' \text{Tr} (i\tau_2\chi^T i\tau_2 \chi)
+\text{Tr} (\mu_{2L}\Delta_L \delta_L +
\mu_{2R}\Delta_R\delta_R).
\label{superpot}
\end{eqnarray}
From the superpotential,  the scalar potential, and the 
CP-even Higgs mass matrix, can be calculated
via a standard procedure. Using the fact that for any Hermetian
matrix its smallest eigenvalue must be smaller than that of its
upper left corner $2 \times 2$ submatrix, we obtain from this
mass matrix the upper bound on the mass of the lightest
Higgs boson in the left-right supersymmetric model:
\begin{equation}
m_{h}^2\leq \frac 12 (g_L^2+g_R^2)(\kappa_1^2 +\kappa_2^2)\cos^2
2\beta,
\label{hm}
\end{equation}
where $g_L$ and $g_R$ are the $SU(2)_L$ and $SU(2)_R$ gauge
couplings, respectively, and $\kappa_1$ and $\kappa_2$ are the 
VEV's of the neutral components of $\Phi(2,2,0)$ and $\chi(2,2,0)$,
respectively, with $\tan\beta = \kappa_2/\kappa_1 $.
We note that the upper bound (\ref{hm}) is not only
independent of the supersymmetry breaking parameters
(as in the case of supersymmetric models based on
$SU(2)_L\times U(1)_Y$),
but it is independent of the
$SU(2)_R$ breaking scale, which, {\it a priori},  can be very large,
and also independent of any  R-parity  breaking vacuum expectation
value. The upper bound is  controlled by
$(\kappa_1^2 + \kappa_2^2)$ and the
dimensionless gauge couplings $(g_L$ and $g_R$) only.
Since the former is essentially fixed by the electroweak scale,
the gauge
couplings $g_L$ and $g_R$ determine the bound (\ref{hm}).
Since the right-handed gauge coupling $g_R$ is not known,
the upper  bound
on the right-hand side of (\ref{hm}) comes from the
requirement~\cite{HPP2}  that the
left-right supersymmetric model remains perturbative
below some scale $\Lambda $. The resulting tree level
upper bound is shown 
in Fig.(3). 
The tree-level bound can be considerably larger
than in MSSM, if the difference between the high scale $\Lambda$
and the intermediate scale $M_R$ is small.
The radiative corrections to the upper bound
from top-stop and bottom-sbottom
sector are sizable and of the same form as in the MSSM.
In Fig.(4), we show the radiatively corrected 
upper bound as a function of
top quark mass in the range $150<m_t<200$ GeV, and compare it with the 
corresponding upper bound in the MSSM.
The upper bound increases with increasing $M_R$ scale, and becomes
less restrictive as this scale is increased.
For $M_R=10$ TeV and $m_{top}=175$ GeV, the bound remains below
155 GeV while for $M_R=10^{10}$ GeV it remains below 175 GeV.
It is seen that the mass limits, except for large $\mu_1,\mu'_1,\mu''_1$,
are somewhat higher in SUSYLR than in the MSSM.

\section{The Neutralino Sector of Supersymmetric Models}

In supersymmetric theories with R-parity conservation,
the lightest neutralino is expected to be the lightest
supersymmetric particle. In MSSM the 
fermionic partners of the  
Higgs bosons mix with
the fermionic partners of the gauge bosons to produce
four neutralino states
$\tilde{\chi}^0_i, ~i = 1, 2, 3, 4,$ and two chargino states
$\tilde{\chi}^{\pm}_i, ~i = 1, 2,$. 
An upper bound on the squared mass of the lightest
neutralino $\chi^0_1$
can be obtained by using the fact that the smallest
eigenvalue of the mass squared matrix of the neutralinos
is smaller than the smallest eigenvalue of its upper
left 2 x 2 submatrix
\begin{equation}
\left[ \begin{array}{cc}
M^2_1+M^2_Z\sin^2\theta_W & -M^2_Z\sin\theta_W\cos\theta_W \\
-M^2_z\sin\theta_W\cos\theta_W & M^2_2 + M^2_Z\cos^2\theta_W
\end{array} \right]
\label{N1}
\end{equation}
thereby resulting in the upper bound \cite{PNPN1}
\begin{equation}
M^2_{\chi^0_1} \le \min(M^2_1+M^2_Z\sin^2\theta_W, M^2_2+M^2_Z\cos^2\theta_W),
\label{N2}
\end{equation}
We note that the  uppwer bound (\ref{N2})
is controlled by, 
in addition
to $M_Z$ and $\theta_{W}$, the soft SUSY breaking gaugino masses, $M_1$ 
and $M_2$. 
This is in contrast
to the Higgs sector of MSSM, where the corresponding bound 
on the (tree level)
mass of the lightest Higgs boson 
is controlled by
$M_Z$, and not by supersymmetry breaking masses.
However, the upper bound can become meaningful in theories
with gauge coupling unification.

We recall that, as a consequence of the renormalization group 
equations (RGEs) satisfied by the gauge couplings and the 
gaugino masses in the MSSM, we have $(\alpha_i = g^2_i/4 \pi,
\alpha_U = g^2_U/4\pi)$,
\begin{equation}
M_1(M_Z)/\alpha_1(M_Z) = M_2(M_Z)/\alpha_2(M_Z) = M_3(M_Z)/\alpha_3(M_Z) =
m_{1/2}/\alpha_U,
\label{N3}
\end{equation}
where $M_{1/2}$ is the common gaugino mass at the grand unification scale,
and $\alpha_U$ is the unified coupling constant. It is important to note
that (\ref{N3}), which is a consequence of one-loop renormalization group
equations, is valid in any grand unified theory irrespective of the
particle content. It reduces the three gaugino mass parameters
to one, which we choose to be the gluino mass
$m_{\tilde{g}}$, which is 
equal to $|M_3|$. This results in an upper bound on the
mass of the lightest neutralino as
a function of the gluino mass:
\begin{equation}
M^2_{\chi^0_1} \le M^2_1 + M^2_Z \sin^2 \theta_W 
\simeq (0.02 m^2_{\tilde{g}} +
1924.5)\mbox{GeV}^2.
\label{N4}
\end{equation}
For a gluino mass of $200$ GeV, the upper bound (\ref{N4}) on 
the mass of the lightest neutralino is $52$ GeV,
which increases to $148$ GeV for a gluino mass of
$1$ TeV. The radiative corrections to the upper
bound can vary between $5\%$ and $20\%$ depending
upon the composition of the lightest neutralino.

Although the  upper bound (\ref{N4}) on the lightest neutralino 
mass has been obtained in the MSSM, a similar upper
bound can be obtained in the more general models based on
the gauge group $SU(2)_L \times U(1)_Y$ with an extended
Higgs sector~\cite{PNPN2}. Numerically the upper bound 
in these extended models
can be typically higher than the one in MSSM.

\section*{Acknowledgements}The author would like to thank the
organizers of the 
Fifth Workshop on Particle Physics
Phenomenology (WHEPP5) for inviting him to the Workshop, and for
the hospitality. This work is supported by
Department of Science and Technology under project
No. SP/S2/K--17/94.

\clearpage
\begin{figure}[htb]
\refstepcounter{figure}
\label{fig1.ps}
\addtocounter{figure}{-1}
\begin{center}
\setlength{\unitlength}{1cm}
\begin{picture}(12,12)
\put(-1,1)
{\mbox{\epsfxsize=12cm\epsffile{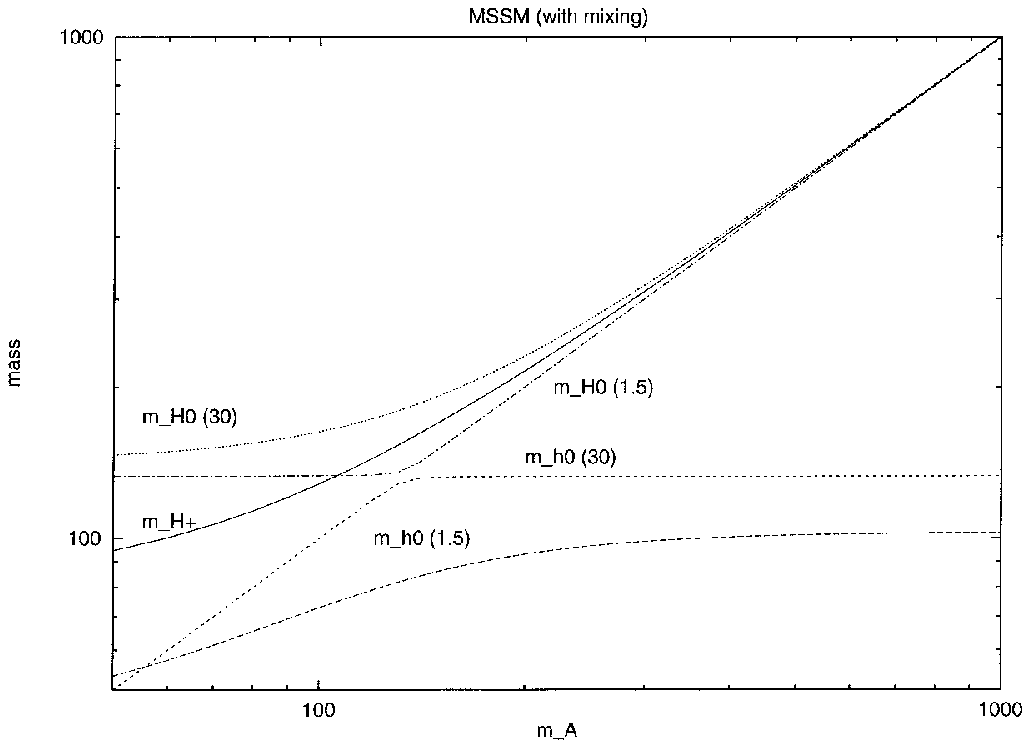}}}
\end{picture}
\vspace*{0mm}
\caption{Masses of the CP-even Higgs bosons $h^0, H^0$
and of the charged Higgs particles $H^{\pm}$ as a
function of the CP-odd Higgs mass $m_A$ for
two values of $\tan\beta = 1.5, 30$.}
\end{center}
\end{figure}
\clearpage
\begin{figure}[htb]
\refstepcounter{figure}
\label{fig2.ps}
\addtocounter{figure}{-1}
\begin{center}
\setlength{\unitlength}{1cm}
\begin{picture}(12,12)
\put(-1,1)
{\mbox{\epsfxsize=12cm\epsffile{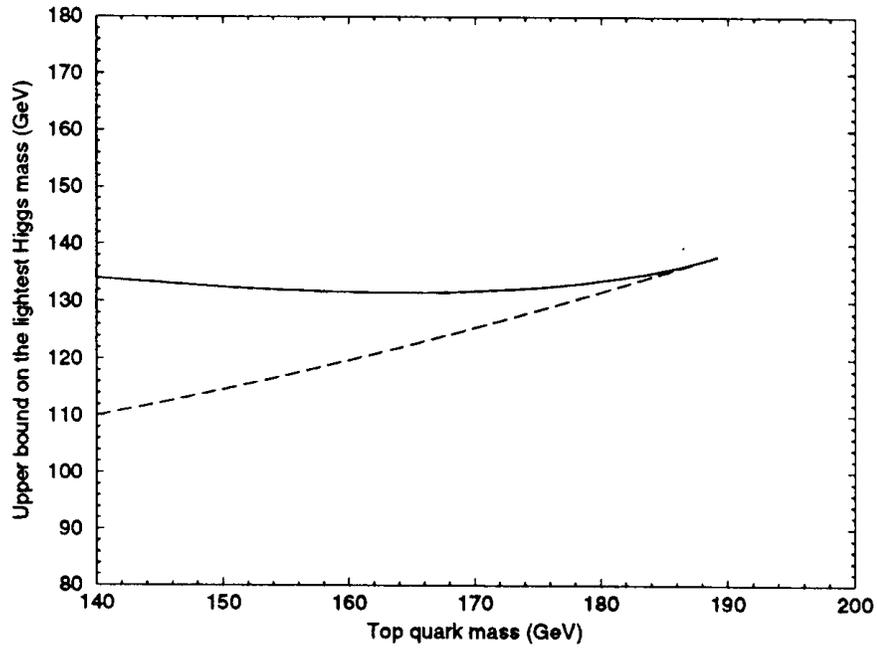}}}
\end{picture}
\vspace*{0mm}
\caption{The upper bound on the mass of the lightest
CP-even Higgs boson in the Non-Minimal Supersymmetric Standard Model
(solid line). We have taken stop mass to be $1$ TeV. The dotted
line shows the corresponding upper bound in the MSSM.}
\end{center}
\end{figure}
\clearpage
\begin{figure}[htb]
\refstepcounter{figure}
\label{fig3.eps}
\addtocounter{figure}{-1}
\begin{center}
\setlength{\unitlength}{1cm}
\begin{picture}(12,12)
\put(-1,1)
{\mbox{\epsfysize=12cm\epsffile{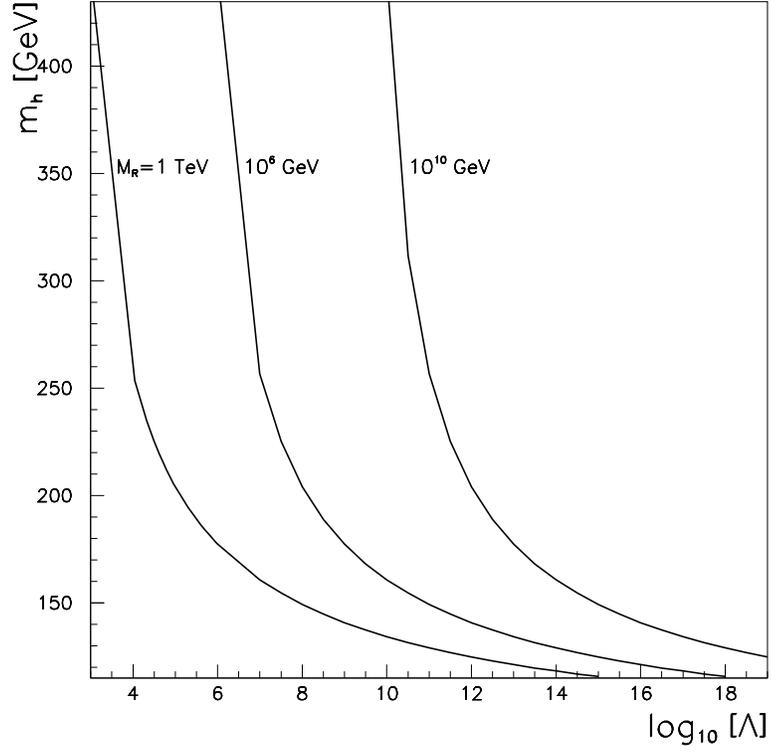}}}
\end{picture}
\vspace*{0mm}
\caption{The tree-level upper bound on the lightest Higgs mass as
a function of the scale $\Lambda$ up to which the $g_R$ coupling
remains perturbative.
The plotted $SU(2)_R\times U(1)_{B-L}$ breaking scales are
$M_R=1$ TeV, $10^6$ GeV and $10^{10}$ GeV.}
\end{center}
\end{figure}
\clearpage
\begin{figure}[htb]
\refstepcounter{figure}
\label{fig4.eps}
\addtocounter{figure}{-1}
\begin{center}
\setlength{\unitlength}{1cm}
\begin{picture}(12,12)
\put(-1,1)
{\mbox{\epsfysize=12cm\epsffile{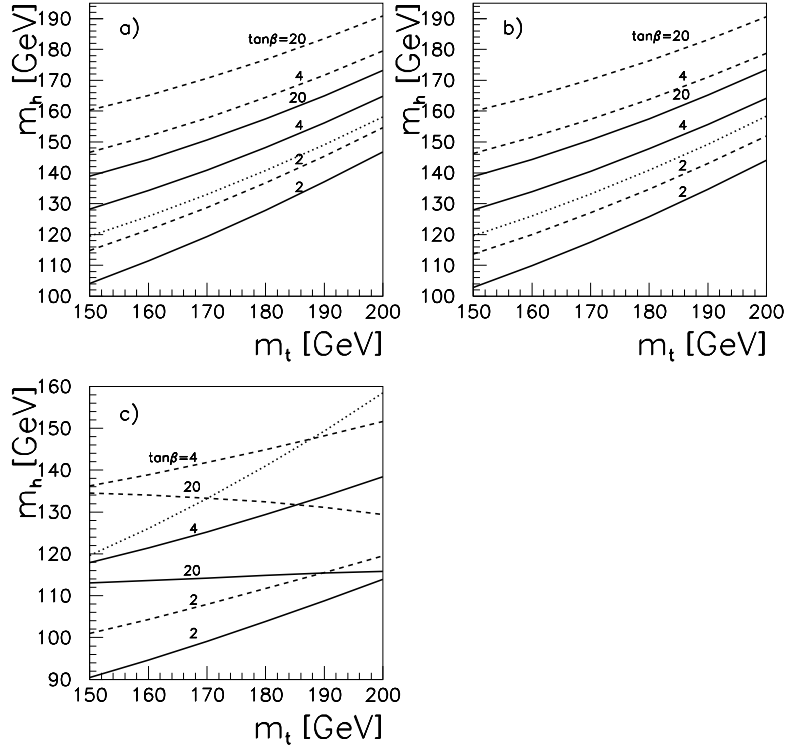}}}
\end{picture}
\vspace*{0mm}
\caption{ The radiatively corrected upper limit on the mass of
the lightest Higgs boson as a
function of $m_t$ with $\Lambda = 10^{16}$ GeV and $A_t=A_b=1$ =
TeV.
The solid line corresponds to the $SU(2)_R$ scale
of 10 TeV and the dashed line to the  $SU(2)_R$ scale
of $10^{10}$ GeV.
The dotted curve corresponds to MSSM limit for $\tan\beta =20$
and $\mu=\mu_1$.
In a)  $\mu_1=\mu'_1=\mu''_1=0$,
in b)  $\mu_1=\mu'_1=\mu''_1=500$ GeV, and
in c) $\mu_1=\mu'_1=\mu''_1=1000$ GeV.}
\end{center}
\end{figure}

\end{document}